# *The lower bound on the energy for bounded systems is equivalent to the Bekenstein upper bound on the entropy to energy ratio for bounded systems*

**Franz-Josef Schmitt**
Berlin Institute of Technology, Strasse des 17. Juni 135, D-10623 Berlin

***Abstract*** *Several approaches can be used to proof the assumption that an universal upper bound on the entropy to energy ratio (S/E) exists in bounded systems. In 1981 Jacob D. Bekenstein published his findings that S/E is limited by the "effective radius" of the system and mentioned various approaches to derive S/E employing quantum statistics or thermodynamics.*
*It can be shown that similar results are obtained considering the energetic difference of longitudinal eigenmodes inside a closed cavity like it was done by Max Planck in 1900 to derive the correct formula for the spectral distribution of the black-body radiation. Considering an information theoretical approach this derivation suggests that the variance of an expectation value $\Delta\langle O \rangle$ is the same like a variance of the probability $\Delta p*$ for measuring $\langle O \rangle$:*
$\Delta\langle O \rangle = \Delta p * \cdot \langle O \rangle$. *Implications of these findings are shortly discussed.*

**Keywords:** Heisenberg´s uncertainty relation, black body radiation, Bekenstein limit

## 1 Introduction

In 1900 Max Planck assumed quantisized portions of energy to explain the radiation spectrum of a cavity on the temperature *T*. Due to Planck the energy of the electromagnetic field is quantisized in portions of $\Delta E = h\nu$. For that reason the probability of high frequency eigenmodes in the cavity with $h\nu \gg k_B T$ becomes proportional to the Boltzman factor $\exp(-h\nu/kT)$. From this finding Max Planck derived the correct formula for the frequency distribution inside the cavity well known as Planck´s law for the black-body radiation [1].

The assumption of Planck that $\Delta E = h\nu$ is in line with the constraint that the amplitude of standing waves inside a cavity vanishes at the surface of the cavity.

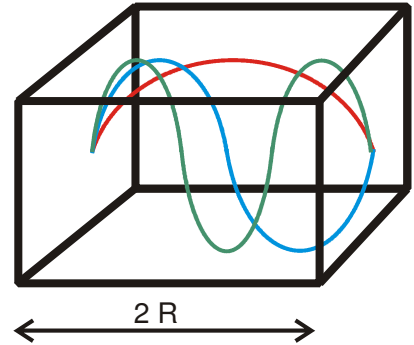

**Fig. 1 Cavity with standing waves**

Due to this constraint the lower bound on the energy in a bounded system depends on its diameter because it is not possible that a quantum state with longer wavelength than twice the radius of a closed cavity might exist inside the bounded system.

$$\lambda_{max} = 4R$$

This leads to

$$E_{min} = h\nu_{min} = \frac{hc}{\lambda_{max}} = \frac{hc}{4R}$$

The energy states in any cavity are discrete because the boundary condition of the cavity allows only states with

$$\lambda = \frac{4R}{N}, N = 1,2,3,...$$

and therefore it follows for the distance between the neighbouring frequencies that

$$\nu_i - \nu_{i-1} = \frac{c}{4R}$$

The energetic distance between two states is

$$\Delta E = E_i - E_{i-1} = \frac{hc}{4R} = h(\nu_i - \nu_{i-1})$$

$$\Delta E = h\nu_{min} = E_{min}$$

The picture of a wavefunction which is subject to the constraints of a bounded system suggests us the quantization of the electromagnetic field in units of $\Delta E = h\nu$ for each mode of the frequency $\nu$. The lowest possible excitation energy of this mode is also $E_{min} = h\nu$.

This assumption of Planck is the basis for quantum theory of the electromagnetic field explaining a huge number of effects like e.g. the photoeffect by Albert Einstein in 1905 [2].

The fact that no particles with $E < \frac{hc}{4R}$ can be found inside a volume with the diameter $R$ can be directly experimentally observed as the so called "Casimir effect" [3].

## 2 The cavity as a measurement device

If we consider the cavity shown in fig. 1 as a measurement device, then the possible outputs $E$ for energy measurements with this device are found to be elements of the discrete series

$$E = n\Delta E = nh\nu_{min} = n\frac{hc}{4R}$$

with an integer $n > 0$.

This fact is independent from the system the measurement is performed on. It is an intrinsic discreetness that follows from the properties of the measurement device only (the cavity shown in fig. 1). Interestingly the cavity radiation is a very common "measurement device" in any case when a simple white light lamp is used as a radiation source of spectroscopic measurements.

As suggested by Caslav Bruker and Anton Zeilinger ([4]) the probabilistic structure of quantum theory should be assumed to be caused by the probabilistic response of a measurement device which is in general not able to represent exactly the eigenvalues of a tested quantum system. If a quantum system is measured the response of a complicated measurement device (e.g. complicated in comparison to 1 bit of information contained in a spin system) would always be necessarily random. This output of the measurement should be considered as an update of information we have about the analyzed system.

Schrödinger called the update of information by a measurement a "special sudden change of the wave function which is different from the smooth behaviour described by the dynamic equation" (the Schrödinger equation) [5]. Schrödinger believed that due to this sudden change the wave function must not be identified with the real object that is observed but it is a mathematical representation of our knowledge about the system.

Schrödinger already mentioned that the understanding of realism in physics might be too strict if one connects realism too strongly with the mathematically constructed wave function. Therefore there is no contradiction between the sudden change of the wave function and causality.

Brukner and Zeilinger suggested that the quantum state carries only a very limited amount of information. The quantum state does not "know" the complex structure of the measurement device and therefore a measurement can not lead to a well defined output. The output is intrinsically random.

Due to this inherent lack of information about the question which state of the measurement device will produce a result (will lead to a "klick" when the measurement is performed) the measurement will lead to a probability distribution along the possible eigenvalues of the measurement device.

In general the measured state should be represented randomly by different eigenstates of the measurement device. For simplicity we assume that only two states $E_i$ and $E_{i+1}$ contribute to the measurement output.

In that case the output of the measurement device would be substantially random at least between the two eigenstates $E_i$ and $E_{i+1}$ with $E_i < E_{system}$ and $E_{i+1} > E_{system}$ leading to an uncertainty which is correlated with the distance of the energetic states of the measurement device.

$$\Delta E_{device} = E_{i+1} - E_i = \frac{hc}{4R}.$$

In the most simple and rough estimation we have an uncertainty which is about the distance between two energy states:

$$\Delta E_{measurement} \approx \Delta E_{device} = \frac{hc}{4R} \qquad (1)$$

This uncertainty is an lower limit for the uncertainty of energy measurements and therefore

$$\Delta E_{measurement} > \frac{hc}{4R} \qquad (2)$$

We find that the uncertainty of energy measurements is in the order of $hc$ and inverse proportional to the "radius" of the measurement device.

## 3 Quantization and Uncertainty

### 3.1 The energy-time uncertainty

Due to the fact that light travels with $c$ ($2R/c = \Delta t$) and needs the time $\Delta t$ to travel across the cavity one gets the uncertainty relation of Heisenberg for energy and time from formula (2) that is

$$\Delta E \Delta t > \frac{h}{2}$$

This fact is well known, of course. Here we can see that in the picture of a radiation field filling a cavity Heisenberg´s uncertainty relation is equivalent to the discreetness of the eigenstates inside the cavity.

A fundamental lower bound on the uncertainty of probability measurements was derived with an information theoretical approach employing the so called Bekenstein limit [6] as suggested by M. Müller [7]. These limitations have implications on the principally achievable resolution of experimental setups, e.g. for time- and space resolved fluorescence spectroscopy [8]. Furthermore, if one accepts that the variance of an expectation value $\Delta\langle O\rangle$ is the same like a variance of the probability $\Delta p*$ for measuring $\langle O\rangle$

$$\Delta\langle O\rangle = \Delta p* \cdot \langle O\rangle \qquad (3)$$

then the general lower bound on probability measurements suggested by M. Mueller leads to the same results like the uncertainty relation of Heisenberg but seems to be more general [8]. This finding will be shortly discussed in chapter 4.

## 3.2 The general uncertainty of probability measurements in bounded systems

M.Müller derived from [6] in 2007 that

$$\Delta p* \geq \frac{h(p*)\hbar c \ln 2}{6\sqrt{2\pi}ER} \qquad (4)$$

as a general uncertainty $\Delta p*$ when measuring unknown probability values $p*$ in a bounded system [7].
$h(p*)$ denotes the binary entropy of the probability $p*$ defined as $h(p*) = -p*\log_2(p*) - (1-p*)\log_2(1-p*)$. For $p* = 0 \vee p* = 1$ the binary entropy $h(p*) = 0$ is minimal and maximal for $p* = 1/2$ with $h(p*) = 1$.
From relation (4) it follows that $\Delta p*$ is in the order of $hc$ and that it is inversely proportional to the system's effective radius and energy. Except the exact numerical value relation (4) follows from relation (2) assuming that $E \cdot \Delta p* = \Delta E$ according to equation (3). Then relation (2) becomes

$$\Delta p* > \frac{hc}{4ER} \qquad (5)$$

(5) is surely fulfilled if (4) is fulfilled.
But (5) can be derived from the black body radiation if $E \cdot \Delta p* = \Delta E$, i.e. the assumption of eq. (3) is correct.
That means that equation (5) is correct without doubts if (3) is correct.
If formula (5) is correct then (4) must be correct ignoring the exact numerical factor $\frac{h(p*)\ln 2}{3\sqrt{2\pi^2}} \approx 1$.
In fact this approch proofes the correctness of the Bekenstein limit in an information theoretical sense.

At this point we should denote again that the relations (4) and (5) are expressions more fundamental than but leading to Heisenberg´s uncertainty relation if one generally accepts that the variance of an expectation value $\Delta\langle O\rangle$ is the same like a variance of the probability $\Delta p*$ for measuring $\langle O\rangle$:
$\Delta\langle O\rangle = \Delta p* \cdot \langle O\rangle$

## 4 Discussion

With (3) the relation (5) can be derived from the eigenmodes of the radiation field in a closed cavity.
But the finding of M. Müller (2007) clearly shows that (5) is a fact of much more general nature like it was formulated by M. Müller in the form of (4) and holds for any probability measured on any bounded system.

Therefore the information theoretical assumption of limited information inside a space volume (Bekenstein) is equivalent to the assumption that the size of a space volume is limiting the existence of particles by reason of the length of the wave traces (no particle can exist in a space volume with a wavelength bigger than twice the diameter of the space).

The generalized form of (4) and (5) suggests that the uncertainty of probability measurements is not a principal limitation of the measurement but an intrinsic uncertainty of any quantum state which already had been formulated by M. Müller when he assumed that "probability gets fuzzy" which is equal to the fact that pure quantum states do not exist on a very small space scale.

It is very interesting to speculate about further implications of this fact, e.g. the fact that the exactness of nature constants can not be infinite.

## 5 Conclusion and further remarks

From relation (5) the well known restrictions due to Heisenberg´s uncertainty relation follow with eq. (3) if $E$ and $R$ are chosen appropriate for a certain experimental setup. $E$ and $R$ might be variables describing an "effective" energy and/or effective radius of the system which might be compared with a product of time and intensity (see also [8]).

High resolved pictures in the fluorescence microscope as presented by Hell et al [9] use the refinement of resolution by a huge statistics of photons.

One could assume that the channel width of such a microscopic detector is a general limitation of the resolution. That means that e.g. in microscopy the pixel number of a CCD will limit the resolution in a general way.

But it has already been shown that it is possible to resolve probability distributions with exactness better than the diameter of a single detection channel if a fit of the distribution over the discrete channels is performed. The position of the fit maximum is not necessarily more uncertain than the channel width.

This principle is illustrated in fig. 2 and already implemented in commercial experimental setups like atomic force microscopes.

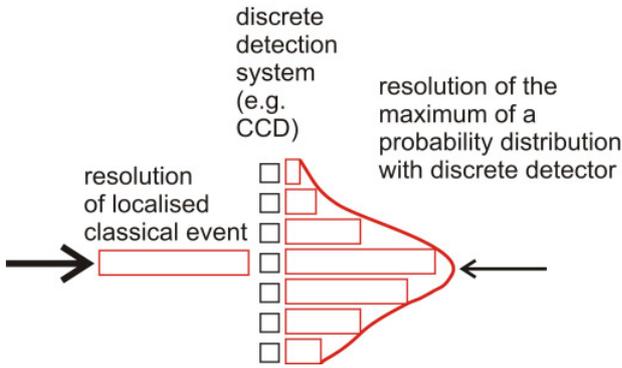

**fig. 2** Resolution enhancement by probability distributions of incoming photons. A broad probability distribution (right side) detected with a discrete medium might be better resolved than a delta-distribution (left side) on the same medium if an appropriate fit is achieved.

The paradox conclusion is that due to the fact that any registration medium is a discrete one (in fact no continuous medium exists) the resolution would be better if the incoming photons are delocalized in a broad probability distribution than if they are localized in a delta-distribution, because the latter one would only hit one channel of the discrete registration medium and could not be resolved below the channel width, while a broad distribution can be fit and the maximum of the probability distribution might be positioned more exact than the width of a channel, if the form of the probability function is known (e.g. a Lorentzian for spatial or energetic distribution).

Of course the question arises if there is any limitation of the achievable resolution if, just assuming a thought experiment, one has an infinite amount of time and/or energy.

Well – also in this academic case there is ! Without the intention for any further conclusion one could funnywise investigate the fundamental limitation of energy distances existing in the universe due to the fact that the universe is limited by it´s diameter and/or it´s age. Therefore the diameter of the universe is not infinite. Then (2) limits the resolution of energy to the value

$$\Delta E > \frac{hc}{4 R_{Univ}} \qquad (6)$$

and due to $R_{Univ} \approx c \cdot T_{Univ}$ one gets the simple relation between energy resolution and the age of the universe:

$$\Delta E \approx \frac{h}{T_{Univ}} \qquad (7)$$

delivering the incredible low value of $\Delta E \approx 10^{-50} J$. Relation (7) is an energy-time-uncertainty according to Heisenberg. It might indicate that the universe itself can not distinguish energy levels in the range $< 10^{-50} J$. Only if the universe expands continuously one would reach an infinite resolution at the end of time.

This general limitation really suggests that there is an inherent uncertainty of nature constants. The finding appears nearly to be trivial if we consider it as the fact that the whole information which is available in the universe is limited. Therefore it is not possible to extract more information from a single experiment than the universe's whole information content

# 6 Acknowledgements

I thank M. Müller for fruitful discussions and the motivation of this paper. His work was the main input for deriving the ideas of this paper.

My thank further belongs to Prof. Hans Joachim Eichler, Prof. Gernot Renger and Dr. Hann-Jörg Eckert supporting my work with helpful information especially in the field of biotechnological research.

This work was supported by the German research foundation DFG in the framework of Sfb 429 "Molecular Physiology, Energetics, and Regulation of Primary Metabolism in Plants" and by BMBF, "optical methods of cleaning and contamination control" (01RI0645).

# 7 References

[1] M. Planck "On the Law of Distribution of Energy in the Normal Spectrum". Annalen der Physik, vol. 4, p. 553 ff (1901)

[2] A. Einstein "On a Heuristic Viewpoint Concerning the Production and Transformation of Light", Annalen der Physik 17: 132–148 (1905)

[3] M.J. Sparnaay, "Measurement of attractive forces between flat plates", Physica 24, 751 (1958)

[4] C. Brukner, A. Zeilinger, "Information and fundamental elements of the structure of quantum theory", arXiv:quant-ph/0212084v1 13 Dec 2002 (Dated: February 1, 2008)

[5] E. Schrödinger, Naturwissenschaften 23, 807-812; 823-828; 844-849. Translation published in Proc. Am. Phil. Soc. 124, p. 323-338 and in Quantum theory and Measurement edited by J.A. Wheeler and W.H. Zurek (Princeton University Press, Princeton), p. 152-167 (1935)

[6] J.D. Bekenstein, "A universal upper bound on the entropy to energy ratio for bounded systems", Phys. Rev. D 23, 287 (1981)

[7] M. Müller, "A lower bound on the uncertainty of probability measurements in closed systems", arXiv:0712.4090v1 [hep-th] (Dec. 2007)

[8] F.-J. Schmitt, "resolution limits of time- and space-correlated single photon counting", In: Proceedings of the 2008 international conference on information theory and statistical learning ITSL 2008, editors: M.Dehmer, M.Drmota, F. Emmert-Streib, ISBN: 1-60132-079-5, pp.91-97 (2008)

[9] S.W. Hell, "Strategy for far-field optical imaging and writing without diffraction limit", Phys.Lett.A 326 (1-2): 140-145 (2004)